
\input phyzzx
%
%
   \message{V 1.18 mods and bug fixes by M.Weinstein}
   \def\unlock{\catcode`@=11}

   \def\lock{\catcode`@=12}

   \unlock
%
%
   \def\PRrefmark#1{~[#1]}
   \def\refitem#1{\ifPhysRev\r@fitem{[#1]}\else\r@fitem{#1.}\fi}
   \def\generatefootsymbol{%
      \ifcase\footsymbolcount%
          \fd@f 13F \or \fd@f 279 \or \fd@f 27A %
            \or \fd@f 278 \or \fd@f 27B %
      \else
         \ifnum\footsymbolcount <0 %
            \xdef\footsymbol{\number-\footsymbolcount}
         \else %
            \fd@f 203
               {\loop \ifnum\footsymbolcount >5
                  \fd@f{203 \footsymbol } %
                  \advance\footsymbolcount by -1%
                \repeat %
               }%
         \fi%
      \fi%
   }
   \def\OldPhysRevRefmark{\let\PRrefmark=\attach}
   \def\OldPRRefitem#1{\r@fitem{#1.}}
   \def\OldPhysRevRefitem{\let\refitem=\OldPRRefitem}
   \def\NPrefs{\let\refmark=\NPrefmark \let\refitem=\NPrefitem}
%
    \newif\iffileexists              \fileexistsfalse
    \newif\ifforwardrefson           \forwardrefsontrue
    \newif\ifamiga                   \amigafalse
    \newif\iflinkedinput             \linkedinputtrue
    \newif\iflinkopen                \linkopenfalse
    \newif\ifcsnameopen              \csnameopenfalse
    \newif\ifdummypictures           \dummypicturesfalse
    \newif\ifcontentson              \contentsonfalse
    \newif\ifcontentsopen            \contentsopenfalse
    \newif\ifmakename                \makenamefalse
    \newif\ifverbdone
    \newif\ifusechapterlabel         \usechapterlabelfalse
    \newif\ifstartofchapter          \startofchapterfalse
    \newif\iftableofplates           \tableofplatesfalse
    \newif\ifplatesopen              \platesopenfalse
    \newif\iftableoftables           \tableoftablesfalse
    \newif\iftableoftablesopen       \tableoftablesopenfalse
    \newif\ifwarncsname              \warncsnamefalse
%
    \newwrite\linkwrite
    \newwrite\csnamewrite
    \newwrite\contentswrite
    \newwrite\plateswrite
    \newwrite\tableoftableswrite
    \newread\testifexists
    \newread\verbinfile

    \newtoks\jobdir                  \jobdir={}
    \newtoks\tempnametoks            \tempnametoks={}
    \newtoks\oldheadline             \oldheadline={}
    \newtoks\oldfootline             \oldfootline={}
    \newtoks\subsectstyle            \subsectstyle={\Number}
    \newtoks\subsubsectstyle         \subsubsectstyle={\Number}
    \newtoks\runningheadlines        \runningheadlines={\relax}
    \newtoks\chapterformat           \chapterformat={\titlestyle}
    \newtoks\sectionformat           \sectionformat={\relax}
    \newtoks\subsectionformat        \subsectionformat={\relax}
    \newtoks\subsubsectionformat     \subsubsectionformat={\relax}
    \newtoks\chapterfontstyle        \chapterfontstyle={\bf}
    \newtoks\sectionfontstyle        \sectionfontstyle={\rm}
    \newtoks\subsectionfontstyle     \subsectionfontstyle={\rm}
    \newtoks\sectionfontstyleb       \sectionfontstyleb={\caps}
    \newtoks\subsubsectionfontstyle  \subsubsectionfontstyle={\rm}

    \newcount\subsectnumber           \subsectnumber=0
    \newcount\subsubsectnumber        \subsubsectnumber=0


   \newdimen\pictureindent           \pictureindent=15pt
   \newdimen\str
   \newdimen\squareht
   \newdimen\squarewd
   \newskip\doublecolskip
   \newskip\tableoftablesskip        \tableoftablesskip=\baselineskip


   \newbox\squarebox


   \newskip\sectionindent            \sectionindent=0pt
   \newskip\subsectionindent         \subsectionindent=0pt
  \def\thechapterhead{\relax}
  \def\thesectionhead{\relax}
  \def\thesubsecthead{\relax}
  \def\thesubsubsecthead{\relax}


   \def\GetIfExists #1 {
       \immediate\openin\testifexists=#1
       \ifeof\testifexists
           \immediate\closein\testifexists
       \else
         \immediate\closein\testifexists
         \input #1
       \fi
   }


   \def\stripbackslash#1#2*{\def\strippedname{#2}}

   \def\ifundefined#1{\expandafter\ifx\csname#1\endcsname\relax}

   \def\val#1{%
      \expandafter\stripbackslash\string#1*%
      \ifundefined{\strippedname}%
      \message{Warning! The control sequence \noexpand#1 is not defined.} ? %
      \else\csname\strippedname\endcsname\fi%
   }
%
%
   \def\CheckForOverWrite#1{%
      \expandafter\stripbackslash\string#1*%
      \ifundefined{\strippedname}%
      \else%
         \ifwarncsname
            \message{Warning! The control sequence \noexpand#1 is being
          overwritten.}%
          \else
          \fi
      \fi%
   }

   \def\FootNoteFonts{\Tenpoint}

   \def\Vfootnote#1{%
      \insert\footins%
      \bgroup%
         \interlinepenalty=\interfootnotelinepenalty%
         \floatingpenalty=20000%
         \singl@true\doubl@false%
         \FootNoteFonts%
         \splittopskip=\ht\strutbox%
         \boxmaxdepth=\dp\strutbox%
         \leftskip=\footindent%
         \rightskip=\z@skip%
         \parindent=0.5%
         \footindent%
         \parfillskip=0pt plus 1fil%
         \spaceskip=\z@skip%
         \xspaceskip=\z@skip%
         \footnotespecial%
         \Textindent{#1}%
         \footstrut%
         \futurelet\next\fo@t%
   }

   \def\csnamech@ck{%
       \ifcsnameopen%
       \else%
           \global\csnameopentrue%
           \immediate\openout\csnamewrite=\the\jobdir\jobname.csnames%
           \immediate\write\csnamewrite{\unlock}%
       \fi%
   }

   \def\linksch@ck{%
          \iflinkopen%
          \else%
              \global\linkopentrue%
              \immediate\openout\linkwrite=\the\jobdir\jobname.links%
          \fi%
   }

   \def\c@ntentscheck{%
       \ifcontentsopen%
       \else%
           \global\contentsopentrue%
           \immediate\openout\contentswrite=\the\jobdir\jobname.contents%
           \immediate\write\contentswrite{%
                \noexpand\titlestyle{Table of Contents}%
           }%
           \immediate\write\contentswrite{\noexpand\bigskip}%
       \fi%
   }

   \def\t@bleofplatescheck{%
       \ifplatesopen%
       \else%
           \global\platesopentrue%
           \immediate\openout\plateswrite=\the\jobdir\jobname.plates%
           \immediate\write\plateswrite{%
                \noexpand\titlestyle{Illustrations}%
           }%
           \immediate\write\plateswrite{%
              \unlock%
           }%
           \immediate\write\plateswrite{\noexpand\bigskip}%
       \fi%
   }

   \def\t@bleoftablescheck{%
       \iftableoftablesopen%
       \else%
           \global\tableoftablesopentrue%
          \immediate\openout\tableoftableswrite=\the\jobdir\jobname.tables%
           \immediate\write\tableoftableswrite{%
                \noexpand\titlestyle{Tables}%
           }%
           \immediate\write\tableoftableswrite{%
              \unlock%
           }%
           \immediate\write\tableoftableswrite{\noexpand\bigskip}%
       \fi%
   }


   \def\linkinput#1 {\input #1
       \iflinkedinput \relax \else \global\linkedinputtrue \fi
       \linksch@ck
       \immediate\write\linkwrite{#1}
   }


   \def\fil@#1 {%
       \ifdummypictures%
          \fileexistsfalse%
          \picfilename={}%
       \else%
          \fileexiststrue%
          \picfilename={#1}%
       \fi%
       \iflinkedinput%
          \iflinkopen \relax%
          \else%
            \global\linkopentrue%
            \immediate\openout\linkwrite=\the\jobdir\jobname.links%
          \fi%
          \immediate\write\linkwrite{#1}%
       \fi%
   }
   \def\Picture#1{%
      \gl@bal\advance\figurecount by 1%
      \CheckForOverWrite#1%
      \xdef#1{\the\figurecount}\pl@t%
      \selfcaptionedtrue%
   }

   \def\s@vepicture{%
       \iffileexists \parsefilename \redopicturebox \fi%
       \ifdim\captionwidth>\z@ \else \captionwidth=\p@cwd \fi%
       \xdef\lastpicture{%
          \iffileexists%
             \setbox0=\hbox{\raise\the\yshift \vbox{%
                \moveright\the\xshift\hbox{\picturedefinition}}%
             }%
          \else%
             \setbox0=\hbox{}%
          \fi
          \ht0=\the\p@cht \wd0=\the\p@cwd \dp0=\the\p@cdp%
          \vbox{\hsize=\the\captionwidth \line{\hss\box0 \hss }%
          \ifcaptioned%
             \vskip\the\captionskip \noexpand\Tenpoint%
             \ifselfcaptioned%
                Figure~\the\figurecount.\enspace%
             \fi%
             \the\Caption%
          \fi }%
       }%
       \iftableofplates%
          \ifplatesopen%
          \else%
             \t@bleofplatescheck%
          \fi%
          \ifselfcaptioned%
             \immediate\write\plateswrite{%
                \noexpand\platetext{%
                \noexpand\item{\rm \the\figurecount .}%
                \the\Caption}{\the\pageno}%
             }%
          \else%
             \immediate\write\plateswrite{%
                \noexpand\platetext{\the\Caption}{\the\pageno}%
             }%
          \fi%
       \fi%
       \endgroup%
   }

   \def\platesout{%
      \ifplatesopen
         \immediate\closeout\plateswrite%
         \global\platesopenfalse%
      \fi%
      \input \jobname.plates%
      \lock%
   }

   \def\platetext#1#2{%
       \hbox to \hsize{\vbox{\hsize=.9\hsize #1}\hfill#2}%
       \vskip \tableoftablesskip \vskip\parskip%
   }


   \def\pres@tpicture{%
       \gl@bal\linesabove=\linesabove
       \s@vepicture
       \setbox\picturebox=\vbox{
       \kern \linesabove\baselineskip \kern 0.3\baselineskip
       \lastpicture \kern 0.3\baselineskip }%
       \dimen@=\p@cht \dimen@i=\dimen@
       \advance\dimen@i by \pagetotal
       \par \ifdim\dimen@i>\pagegoal \vfil\break \fi
       \dimen@ii=\hsize
       \advance\dimen@ii by -\pictureindent \advance\dimen@ii by -\p@cwd
       \setbox0=\vbox to\z@{\kern-\baselineskip \unvbox\picturebox \vss }
   }

   \def\subspaces@t#1:#2;{%
      \baselineskip = \normalbaselineskip%
      \multiply\baselineskip by #1 \divide\baselineskip by #2%
      \lineskip = \normallineskip%
      \multiply\lineskip by #1 \divide\lineskip by #2%
      \lineskiplimit = \normallineskiplimit%
      \multiply\lineskiplimit by #1 \divide\lineskiplimit by #2%
      \parskip = \normalparskip%
      \multiply\parskip by #1 \divide\parskip by #2%
      \abovedisplayskip = \normaldisplayskip%
      \multiply\abovedisplayskip by #1 \divide\abovedisplayskip by #2%
      \belowdisplayskip = \abovedisplayskip%
      \abovedisplayshortskip = \normaldispshortskip%
      \multiply\abovedisplayshortskip by #1%
        \divide\abovedisplayshortskip by #2%
      \belowdisplayshortskip = \abovedisplayshortskip%
      \advance\belowdisplayshortskip by \belowdisplayskip%
      \divide\belowdisplayshortskip by 2%
      \smallskipamount = \skipregister%
      \multiply\smallskipamount by #1 \divide\smallskipamount by #2%
      \medskipamount = \smallskipamount \multiply\medskipamount by 2%
      \bigskipamount = \smallskipamount \multiply\bigskipamount by 4%
   }


   \def\makename#1{
       \global\makenametrue
       \global\tempnametoks={#1}
   }

   \def\nomakename#1{\relax}


   \def\savename#1{%
      \CheckForOverWrite{#1}%
      \csnamech@ck%
      \immediate\write\csnamewrite{\def\the\tempnametoks{#1}}%
   }

   \def\FootNoteFonts{\Tenpoint}

   \def\Vfootnote#1{%
      \insert\footins%
      \bgroup%
         \interlinepenalty=\interfootnotelinepenalty%
         \floatingpenalty=20000%
         \singl@true\doubl@false%
         \FootNoteFonts%
         \splittopskip=\ht\strutbox%
         \boxmaxdepth=\dp\strutbox%
         \leftskip=\footindent%
         \rightskip=\z@skip%
         \parindent=0.5%
         \footindent%
         \parfillskip=0pt plus 1fil%
         \spaceskip=\z@skip%
         \xspaceskip=\z@skip%
         \footnotespecial%
         \Textindent{#1}%
         \footstrut%
         \futurelet\next\fo@t%
   }
%

   \def\eqname#1{%
      \CheckForOverWrite{#1}%
      \rel@x{\pr@tect%
      \csnamech@ck%
      \ifnum\equanumber<0%
          \xdef#1{{\noexpand\f@m0(\number-\equanumber)}}%
          \immediate\write\csnamewrite{%
            \def\noexpand#1{\noexpand\f@m0 (\number-\equanumber)}}%
          \gl@bal\advance\equanumber by -1%
      \else%
          \gl@bal\advance\equanumber by 1%
          \ifusechapterlabel%
            \xdef#1{{\noexpand\f@m0(\ifcn@@ \chapterlabel.\fi%
               \number\equanumber)}%
            }%
          \else%
             \xdef#1{{\noexpand\f@m0(\ifcn@@%
                 {\the\chapterstyle{\the\chapternumber}}.\fi%
                 \number\equanumber)}}%
          \fi%
          \ifcn@@%
             \ifusechapterlabel
                \immediate\write\csnamewrite{\def\noexpand#1{(%
                  {\chapterlabel}.%
                  \number\equanumber)}%
                }%
             \else
                \immediate\write\csnamewrite{\def\noexpand#1{(%
                  {\the\chapterstyle{\the\chapternumber}}.%
                  \number\equanumber)}%
                }%
             \fi%
          \else%
              \immediate\write\csnamewrite{\def\noexpand#1{(%
                  \number\equanumber)}}%
          \fi%
      \fi}%
      #1%
   }

   \def\eqn{\eqno\eqname}

   \let\eqnalign=\eqname


   \def\REFNUM#1{%
      \CheckForOverWrite{#1} %
      \rel@x\gl@bal\advance\referencecount by 1%
      \xdef#1{\the\referencecount}%
      \csnamech@ck%
      \immediate\write\csnamewrite{\def\noexpand#1{\the\referencecount}}%
   }

   %

   \def\FIGNUM#1{
      \CheckForOverWrite{#1}%
      \rel@x\gl@bal\advance\figurecount by 1%
      \xdef#1{\the\figurecount}%
      \csnamech@ck%
      \immediate\write\csnamewrite{\def\noexpand#1{\the\figurecount}}%
   }


   \def\TABNUM#1{%
      \CheckForOverWrite{#1}%
      \rel@x \gl@bal\advance\tablecount by 1%
      \xdef#1{\the\tablecount}%
      \csnamech@ck%
      \immediate\write\csnamewrite{\def\noexpand#1{\the\tablecount}}%
   }


   \def\tableoftableson{%
      \global\tableoftablestrue%

      \gdef\TABLE##1##2{%
         \t@bleoftablescheck%
         \TABNUM ##1%
         \immediate\write\tableoftableswrite{%
            \noexpand\tableoftablestext{%
            \noexpand\item{\rm \the\tablecount .}%
                ##2}{\the\pageno}%
             }%
      }

      \gdef\Table##1{\TABLE\?{##1}Table~\?}
   }

   \def\tableoftablestext#1#2{%
       \hbox to \hsize{\vbox{\hsize=.9\hsize #1}\hfill#2}%
       \vskip \tableoftablesskip%
   }

   \def\tableoftablesout{%
      \iftableoftablesopen
         \immediate\closeout\tableoftableswrite%
         \global\tableoftablesopenfalse%
      \fi%
      \input \jobname.tables%
      \lock%
   }

%
%
%
%
%
%

   \def\contentsoff{\contentsonfalse}

   \def\f@m#1{\f@ntkey=#1\fam=\f@ntkey\the\textfont\f@ntkey\rel@x}
   \def\em@{\rel@x%
      \ifnum\f@ntkey=0\it%
      \else%
         \ifnum\f@ntkey=\bffam\it%
         \else\rm  %
         \fi%
      \fi%
   }

   \def\fontsoff{%
      \def\mit{\relax}%
      \let\oldstyle=\mit%
      \def\cal{\relax}%
      \def\it{\relax}%
      \def\sl{\relax}%
      \def\bf{\relax}%
      \def\tt{\relax}%
      \def\caps{\relax}%
      \let\cp=\caps%
   }


   \def\fontson{%
      \def\rm{\n@expand\f@m0}%
      \def\mit{\n@expand\f@m1}%
      \let\oldstyle=\mit%
      \def\cal{\n@expand\f@m2}%
      \def\it{\n@expand\f@m\itfam}%
      \def\sl{\n@expand\f@m\slfam}%
      \def\bf{\n@expand\f@m\bffam}%
      \def\tt{\n@expand\f@m\ttfam}%
      \def\caps{\n@expand\f@m\cpfam}%
      \let\cp=\caps%
   }

   \fontson
%


   \def\@alpha#1{\count255='140 \advance\count255 by #1\char\count255}
   \def\alphabetic{\@alpha}
   \def\@Alpha#1{\count255='100 \advance\count255 by #1\char\count255}
   \def\Alphabetic{\@Alpha}
   \def\@Roman#1{\uppercase\expandafter{\romannumeral #1}}
   \def\Roman{\@Roman}
   \def\@roman#1{\romannumeral #1}
   \def\roman{\@roman}
   \def\@number#1{\number #1}
   \def\Number{\@number}

   \def\leaderfill{\leaders\hbox to 1em{\hss.\hss}\hfill}

   \def\chapterinfo#1{%
      \line{%
         \ifcn@@%
            \hbox to \itemsize{\hfil\chapterlabel .\quad\ }%
         \fi%
         \noexpand{#1}\leaderfill\the\pagenumber%
      }%
   }

   \def\sectioninfo#1{%
      \line{%
         \ifcn@@%
            \hbox to 2\itemsize{\hfil\sectlabel \quad}%
          \else%
            \hbox to \itemsize{\hfil\quad}%
          \fi%
          \ \noexpand{#1}%
          \leaderfill \the\pagenumber%
      }%
   }

   \def\subsectioninfo#1{%
      \line{%
         \ifcn@@%
            \hbox to 3\itemsize{\hfil \quad\subsectlabel\quad}%
         \else%
            \hbox to 2\itemsize{\hfil\quad}%
         \fi%
          \ \noexpand{#1}%
          \leaderfill \the\pagenumber%
      }%
   }

   \def\subsubsecinfo#1{%
      \line{%
         \ifcn@@%
            \hbox to 4\itemsize{\hfil\subsubsectlabel\quad}%
         \else%
            \hbox to 3\itemsize{\hfil\quad}%
         \fi%
         \ \noexpand{#1}\leaderfill \the\pagenumber%
      }%
   }

   \def\CONTENTS#1;#2{
       {\let\makename=\nomakename
        \if#1C
            \immediate\write\contentswrite{\chapterinfo{#2}}%
        \else\if#1S
                \immediate\write\contentswrite{\sectioninfo{#2}}%
             \else\if#1s
                     \immediate\write\contentswrite{\subsectioninfo{#2}}%
                  \else\if#1x
                          \immediate\write\contentswrite{%
                              \subsubsecinfo{#2}}%
                       \fi
                  \fi
             \fi
        \fi
       }
   }

   \def\chapterreset{\gl@bal\advance\chapternumber by 1%
       \ifnum\equanumber<0 \else\gl@bal\equanumber=0 \fi%
       \gl@bal\sectionnumber=0  \gl@bal\let\sectlabel=\rel@x%
       \gl@bal\subsectnumber=0   \gl@bal\let\subsectlabel=\rel@x%
       \gl@bal\subsubsectnumber=0 \gl@bal\let\subsubsectlabel=\rel@x%
       \ifcn@%
           \gl@bal\cn@@true {\pr@tect\xdef\chapterlabel{%
           {\the\chapterstyle{\the\chapternumber}}}}%
       \else%
           \gl@bal\cn@@false \gdef\chapterlabel{\rel@x}%
       \fi%
       \gl@bal\startofchaptertrue%
   }

   \def\chapter#1{\par \penalty-300 \vskip\chapterskip%
       \spacecheck\chapterminspace%
       \gdef\thechapterhead{#1}%
       \gdef\thesectionhead{\relax}%
       \gdef\thesubsecthead{\relax}%
       \gdef\thesubsubsecthead{\relax}%
       \chapterreset \the\chapterformat{\the\chapterfontstyle%
          \ifcn@@\chapterlabel.~~\fi #1}%
       \nobreak\vskip\headskip \penalty 30000%
       {\pr@tect\wlog{\string\chapter\space \chapterlabel}}%
       \ifmakename%
           \csnamech@ck
           \ifcn@@%
              \immediate\write\csnamewrite{\def\the\tempnametoks{%
                 {\the\chapterstyle{\the\chapternumber}}}%
              }%
            \fi%
            \global\makenamefalse%
       \fi%
       \ifcontentson%
          \c@ntentscheck%
          \CONTENTS{C};{#1}%
       \fi%
       }%

   \def\section#1{\par \ifnum\lastpenalty=30000\else%
       \penalty-200\vskip\sectionskip \spacecheck\sectionminspace\fi%
       \gl@bal\advance\sectionnumber by 1%
       \gl@bal\subsectnumber=0%
       \gl@bal\let\subsectlabel=\rel@x%
       \gl@bal\subsubsectnumber=0%
       \gl@bal\let\subsubsectlabel=\rel@x%
       \gdef\thesectionhead{#1}%
       \gdef\thesubsecthead{\relax}%
       \gdef\thesubsubsecthead{\relax}%
       {\pr@tect\xdef\sectlabel{\ifcn@@%
          {\the\chapterstyle{\the\chapternumber}}.%
          {\the\sectionstyle{\the\sectionnumber}}\fi}%
       \wlog{\string\section\space \sectlabel}}%
       \the\sectionformat{\noindent\the\sectionfontstyle%
            {\ifcn@@\unskip\hskip\sectionindent\sectlabel~~\fi%
                \the\sectionfontstyleb#1}}%
       \par%
       \nobreak\vskip\headskip \penalty 30000%
       \ifmakename%
           \csnamech@ck%
           \ifcn@@%
              \immediate\write\csnamewrite{\def\the\tempnametoks{%
                 {\the\chapterstyle{\the\chapternumber}.%
                  \the\sectionstyle{\the\sectionnumber}}}
              }%
            \fi%
            \global\makenamefalse%
       \fi%
       \ifcontentson%
          \c@ntentscheck%
          \CONTENTS{S};{#1}%
       \fi%
   }

   \def\subsection#1{\par \ifnum\lastpenalty=30000\else%
       \penalty-200\vskip\sectionskip \spacecheck\sectionminspace\fi%
       \gl@bal\advance\subsectnumber by 1%
       \gl@bal\subsubsectnumber=0%
       \gl@bal\let\subsubsectlabel=\rel@x%
       \gdef\thesubsecthead{#1}%
       \gdef\thesubsubsecthead{\relax}%
       {\pr@tect\xdef\subsectlabel{\the\subsectionfontstyle%
           \ifcn@@{\the\chapterstyle{\the\chapternumber}}.%
           {\the\sectionstyle{\the\sectionnumber}}.%
           {\the\subsectstyle{\the\subsectnumber}}\fi}%
           \wlog{\string\section\space \subsectlabel}%
       }%
       \the\subsectionformat{\noindent\the\subsectionfontstyle%
         {\ifcn@@\unskip\hskip\subsectionindent%
          \subsectlabel~~\fi#1}}%
       \par%
       \nobreak\vskip\headskip \penalty 30000%
       \ifmakename%
           \csnamech@ck%
           \ifcn@@%
              \immediate\write\csnamewrite{\def\the\tempnametoks{%
                 {\the\chapterstyle{\the\chapternumber}}.%
                 {\the\sectionstyle{\the\sectionnumber}}.%
                 {\the\subsectstyle{\the\subsectnumber}}}%
              }%
            \fi%
            \global\makenamefalse%
       \fi%
       \ifcontentson%
          \c@ntentscheck%
          \CONTENTS{s};{#1}%
       \fi%
   }

   \def\subsubsection#1{\par \ifnum\lastpenalty=30000\else%
       \penalty-200\vskip\sectionskip \spacecheck\sectionminspace\fi%
       \gl@bal\advance\subsubsectnumber by 1%
       \gdef\thesubsubsecthead{#1}%
       {\pr@tect\xdef\subsubsectlabel{\the\subsubsectionfontstyle\ifcn@@%
           {\the\chapterstyle{\the\chapternumber}}.%
           {\the\sectionstyle{\the\sectionnumber}}.%
           {\the\subsectstyle{\the\subsectnumber}}.%
           {\the\subsubsectstyle{\the\subsubsectnumber}}\fi}%
           \wlog{\string\section\space \subsubsectlabel}%
       }%
       \the\subsubsectionformat{\the\subsubsectionfontstyle%
          \noindent{\ifcn@@\unskip\hskip\subsectionindent%
            \subsubsectlabel~~\fi#1}}%
       \par%
       \nobreak\vskip\headskip \penalty 30000%
       \ifmakename%
           \csnamech@ck%
           \ifcn@@%
              \immediate\write\csnamewrite{\def\the\tempnametoks{%
                {\the\chapterstyle{\the\chapternumber}.%
                 \the\sectionstyle{\the\sectionnumber}.%
                 \the\subsectionstyle{\the\subsectnumber}.%
                 \the\subsubsectstyle{\the\subsubsectnumber}}}%
              }%
            \fi%
            \global\makenamefalse%
       \fi%
       \ifcontentson%
          \c@ntentscheck%
          \CONTENTS{x};{#1}%
       \fi%
   }%

   \def\contentsinput{%
       \ifcontentson%
           \contentsopenfalse%
           \immediate\closeout\contentswrite%
           \global\oldheadline=\headline%
           \global\headline={\hfill}%
           \global\oldfootline=\footline%
           \global\footline={\hfill}%
           \fontsoff \unlock%
           \input \the\jobdir\jobname.contents%
           \fontson%
           \lock%
           \endpage%
           \global\headline=\oldheadline%
           \global\footline=\oldfootline%
       \else%
           \relax%
       \fi%
   }


       \def\phyzzxfootline{
           \footline={\ifletterstyle\the\letterfootline%
               \else\the\paperfootline\fi}%
       }

%

   {\obeyspaces}

   \def\verbfile#1{
       {\catcode`\\=12\catcode`\{=12
       \catcode`\}=12\catcode`\$=12\catcode`\&=12
       \catcode`\#=12\catcode`\%=12\catcode`\~=12
       \catcode`\_=12\catcode`\^=12\obeyspaces\obeylines\tt
       \verbdonetrue\openin\verbinfile=#1
       \loop\read\verbinfile to \inline
           \ifeof\verbinfile
               \verbdonefalse
           \else
              \leftline{\inline}
           \fi
       \ifverbdone\repeat
       \closein\verbinfile}
   }

   \def\boxit#1{\vbox{\hrule\hbox{\vrule\kern3pt%
       \vbox{\kern3pt#1\kern3pt}\kern3pt\vrule}\hrule}%
   }

   \def\square{%
      \setbox\squarebox=\boxit{\hbox{\phantom{x}}}
      \squareht = 1\ht\squarebox
      \squarewd = 1\wd\squarebox
      \vbox to 0pt{
          \offinterlineskip \kern -.9\squareht
          \hbox{\copy\squarebox \vrule width .2\squarewd height .8\squareht
              depth 0pt \hfill
          }
          \hbox{\kern .2\squarewd\vbox{%
            \hrule height .2\squarewd width \squarewd}
          }
          \vss
      }
   }

   \def\fboxit#1#2{
       \vbox{\hrule height #1
           \hbox{\vrule width #1
               \kern3pt \vbox{\kern3pt#2\kern3pt}\kern3pt \vrule width #1
           }
           \hrule height #1
       }
   }

   \let\eqnameold=\eqname

   \def\draft{\def\eqname##1{\eqnameold##1:{\tt\string##1}}
      \let\eqnalign = \eqname
   }
%
%
   \def\runningrightheadline{%
       \hfill%
       \tenit%
       \ifstartofchapter%
          \global\startofchapterfalse%
       \else%
          \ifcn@@ \the\chapternumber.\the\sectionnumber\quad\fi%
              {\fontsoff\thesectionhead}%
       \fi%
       \qquad\twelverm\folio%
   }

   \def\runningleftheadline{%
      \twelverm\folio\qquad%
      \tenit%
      \ifstartofchapter%
          \global\startofchapterfalse%
      \else%
         \ifcn@@%
             Chapter \the\chapternumber \quad%
         \fi%
         {\fontsoff\thechapterhead}%
         \hfill%
      \fi%
   }

   \runningheadlines={%
      \ifodd\pageno%
         \runningrightheadline%
      \else%
         \runningleftheadline%
      \fi
   }

%
%
%
%
%

   \font\dfont=cmr10 scaled \magstep5


   \newbox\cstrutbox
   \newbox\dlbox
   \newbox\vsk

   \setbox\cstrutbox=\hbox{\vrule height10.5pt depth3.5pt width\z@}

   \def\cstrut{\relax\ifmmode\copy\cstrutbox\else\unhcopy\cstrutbox\fi}

   \def\dl #1{\noindent\strut
       \setbox\dlbox=\hbox{\dfont #1\kern 2pt}%
       \setbox\vsk=\hbox{(}%
       \hangindent=1.1\wd\dlbox
       \hangafter=-2
       \strut\hbox to 0pt{\hss\vbox to 0pt{%
         \vskip-.75\ht\vsk\box\dlbox\vss}}%
       \noindent
   }

%
%

   \newdimen\fullhsize

   \fullhsize=6.5in
   \def\fullline{\hbox to\fullhsize}
   \let\l@r=L

   \newbox\leftcolumn
   \newbox\midcolumn

   \def\twocols{ \hsize = 3.1in
%
%
%
%

      \doublecolskip=.3333em plus .3333em minus .1em
      \global\spaceskip=\doublecolskip%
      \global\hyphenpenalty=0

      \singlespace

      \gdef\makeheadline{
          \vbox to 0pt{ \skip@=\topskip
          \advance\skip@ by -12pt \advance\skip@ by -2\normalbaselineskip
          \vskip\skip@
          \fullline{\vbox to 12pt{}\the\headline} \vss}\nointerlineskip
      }

      \def\makefootline{\baselineskip = 1.5\normalbaselineskip
           \fullline{\the\footline}
      }

      \output={
          \if L\l@r
             \global\setbox\leftcolumn=\columnbox \global\let\l@r=R
          \else
              \doubleformat \global\let\l@r=L
          \fi
          \ifnum\outputpenalty>-20000 \else\dosupereject\fi
      }

      \def\doubleformat{
          \shipout\vbox{
             \makeheadline
             \fullline{\box\leftcolumn\hfil\columnbox}
             \makefootline
          }
          \advancepageno
      }

      \def\columnbox{\leftline{\pagebody}}

      \outer\def\twobye{
          \par\vfill\supereject\if R\l@r \null\vfill\eject\fi\end
      }
   }

   \def\threecols{
       \hsize = 2.0in \tenpoint

      \doublecolskip=.3333em plus .3333em minus .1em
      \global\spaceskip=\doublecolskip%
      \global\hyphenpenalty=0

       \singlespace

       \def\makeheadline{\vbox to 0pt{ \skip@=\topskip
           \advance\skip@ by -12pt \advance\skip@ by -2\normalbaselineskip
           \vskip\skip@ \fullline{\vbox to 12pt{}\the\headline} \vss
           }\nointerlineskip
       }
       \def\makefootline{\baselineskip = 1.5\normalbaselineskip
                 \fullline{\the\footline}
       }

       \output={
          \if L\l@r
             \global\setbox\leftcolumn=\columnbox \global\let\l@r=M
          \else \if M\l@r
                   \global\setbox\midcolumn=\columnbox
                   \global\let\l@r=R
                \else \tripleformat \global\let\l@r=L
                \fi
          \fi
          \ifnum\outputpenalty>-20000 \else\dosupereject\fi
       }

       \def\tripleformat{
           \shipout\vbox{
               \makeheadline
               \fullline{\box\leftcolumn\hfil\box\midcolumn\hfil\columnbox}
               \makefootline
           }
           \advancepageno
       }

       \def\columnbox{\leftline{\pagebody}}

       \outer\def\threebye{
           \par\vfill\supereject
           \if R\l@r \null\vfill\eject\fi
           \end
       }
   }


%
%
%

   \everyjob{%

       \immediate\openin\testifexists=m
       \ifeof\testifexists
           \immediate\closein\testifexists
       \else
         \immediate\closein\testifexists
         \input m
       \fi
   yphyx.tex
      \ifforwardrefson%
         
       \immediate\openin\testifexists=\the
       \ifeof\testifexists
           \immediate\closein\testifexists
       \else
         \immediate\closein\testifexists
         \input \the
       \fi
   \jobdir\jobname.csnames
      \fi%
   }

\contentsoff
\lock



\let\lettertopfil=\lettertopskip




\def\l"{``} 
\def\linebreak{\unskip\break}

%
%

\def\ifmath#1{\relax\ifmmode #1\else $#1$\fi}

\def\unlock{\catcode`@=11} 
\def\lock{\catcode`@=12} 



%

%

%

%

\def\uss#1{\vtop{\hbox{#1}\kern 3pt \hrule}} 
\def\~{\accent'24 } 

\def\ls#1{\ifmath{_{\lower1.5pt\hbox{$\scriptstyle #1$}}}}

\def\eg{{\it e.g., }}%
\def\ie{{\it i.e., }}%

\def\SLACHEAD{\setbox0=\vbox{\baselineskip=12pt
      \ialign{\tenfib ##\hfil\cr
         \hskip -17pt\tenit Mail Address:\ \ Bin 81\cr
         SLAC, P.O.Box 4349\cr Stanford, California, 94305\cr}}
   \setbox2=\vbox to\ht0{\vfil\hbox{\caps Stanford Linear Accelerator
         Center}\vfil}
   \smallskip \line{\hskip -7pt\box2\hfil\box0}\bigskip}
%
%


\newbox\figbox
\newdimen\zero  \zero=0pt
\newdimen\figmove
\newdimen\figwidth
\newdimen\figheight
\newdimen\figrefheight
\newdimen\textwidth
\newtoks\figtoks
\newcount\figcounta
\newcount\figlines
\def\figreset{\global\figmove=\baselineskip \global\figcounta=0
\global\figlines=1 \global\figtoks={ } }


\def\picture#1by#2:#3{\global\setbox\figbox=\vbox{\vskip #1
\hbox{\vbox{\hsize=#2 \noindent #3}}}
\global\setbox\figbox=\vbox{\kern 10pt
\hbox{\kern 5pt \box\figbox \kern 10pt }\kern 10pt}
\global\figwidth=1\wd\figbox
\global\figheight=1\ht\figbox
\global\figrefheight=\figheight
\global\textwidth=\hsize
\global\advance\textwidth by - \figwidth }
\def\figtoksappend{\edef\temp##1{\global\figtoks=%
{\the\figtoks ##1}}\temp}
\def\figparmsa#1{\loop \global\advance\figcounta by 1%
\ifnum \figcounta < #1 \figtoksappend{0pt \the\hsize}%
\global\advance\figlines by 1 \repeat }
\newdimen\figstep
\def\figst@p{\global\figstep = \baselineskip}
\def\figparmsb{\loop \ifdim\figrefheight > 0pt%
\figtoksappend{ \the\figwidth \the\textwidth}%
\global\advance\figrefheight by -\figstep%
\global\advance\figlines by 1%
\repeat }


\def\figtext#1:#2{\figreset \figst@p%
\figparmsa{#1}%
\figparmsb%
\multiply\figmove by #1%
\global\setbox\figbox=\vbox to 0pt{\vskip\figmove\hbox{\box\figbox}\vss}%
\parshape=\the\figlines\the\figtoks\the\zero\the\hsize%
\noindent\rlap{\box\figbox} #2}

\def\lozenge{\boxit{\hbox to 1.5pt{\vrule height 1pt width 0pt \hfill}}}

\def\endpoint{\linebreak \noindent\hangindent 0pt after 0 }

\def\cornerarrow{%
   \rlap{ 
      \raise 7pt
      \hbox{\vrule height 6.5pt depth 4pt}%
      }%
   \hskip 0pt plus 10000pt \rightarrow
   }%
%
%

\def\9{\hskip .5em}

\def\|{\vrule height 16pt depth 6 pt}

\def\spose#1{\raise6pt\hbox to 0pt{#1 \hskip0pt minus10000pt}}
\def\.{\hskip -4pt plus 10000pt}

\def\leftline#1{\line{#1\hss}}

\def\rightline#1{\line{\hss#1}}

\def\str{\penalty-10000\hfilneg\ }

\def\mybox#1{\hbox to size{\vbox{\hrule\hbox{\vrule\hskip4pt
\vbox{\vskip4pt #1 \vskip4pt}\hskip4pt\vrule}\hrule}}}

\def\lead{\leaders\hbox to 10pt{\hfill.\hfill}\hfill}

\chapterminspace=12pc
\referenceminspace=12pc
\def\refout{\par\penalty-400\vskip\chapterskip
   \spacecheck\referenceminspace
   \ifreferenceopen \Closeout\referencewrite \referenceopenfalse \fi
   \line{\hfil REFERENCES\hfil}\vskip\headskip
   \input \jobname.refs
   }
%
%
\def\PRL#1&#2&#3&{\sl Phys.~Rev.~Lett.\ \bf #1\rm ,\ #2\ (19#3)}
\def\PRB#1&#2&#3&{\sl Phys.~Rev.\ \bf #1\rm ,\ #2\ (19#3)}
\def\NPB#1&#2&#3&{\sl Nucl.~Phys.\ \bf #1\rm ,\ #2\ (19#3)}
\def\PL#1&#2&#3&{\sl Phys.~Lett.\ \bf #1\rm ,\ #2\ (19#3)}
\def\ZP#1&#2&#3&{\sl Z.~Phys.\ \bf #1\rm ,\ #2\ (19#3)}
\def\NIM#1&#2&#3&{\sl Nucl.~Instrum.~\& Methods\ \bf #1\rm ,\ #2\ (19#3)}
\def\RMP#1&#2&#3&{\sl Rev.~Mod.~Phys.\ \bf #1\rm ,\ #2\ (19#3)}
\def\IEEE#1&#2&#3&{\sl IEEE Trans.~Nucl.~Phys. \bf #1\rm ,\ #2\ (19#3)}
\def\NPRL#1&#2&#3&{\sl Phys.~Rev.~Lett.\ \bf #1\ \rm (19#2)\ #3}
\def\NPR#1&#2&#3&{\sl Phys.~Rev.\ \bf #1\ \rm (19#2)\ #3}
\def\NNP#1&#2&#3&{\sl Nucl.~Phys.\ \bf #1\ \rm (19#2)\ #3}
\def\NPL#1&#2&#3&{\sl Phys.~Lett.\ \bf #1\ \rm (19#2)\ #3}
\def\NZP#1&#2&#3&{\sl Z.~Phys.\ \bf #1\ \rm (19#2)\ #3}
\def\AJ#1&#2&#3&{\sl Ap.~J.\ \bf #1\ \rm (19#2)\ #3}
%
%
\def\CVaddresses#1#2{\hbox to\hsize{
Home:\quad\vtop{\halign{##\hfil\crcr #1}}\hfill
Work:\quad\vtop{\halign{##\hfil\crcr #2}}}}
%
%
\def\pmb#1{\setbox0=\hbox{$#1$}%
     \kern-.025em\copy0\kern-\wd0
     \kern.05em\copy0\kern-\wd0
     \kern-.025em\raise.0433em\box0 }

\def\spmb#1{\setbox0=\hbox{$\tenpoint #1$}%
     \kern-.020em\copy0\kern-\wd0
     \kern.03em\copy0\kern-\wd0
     \kern-.020em\raise.0433em\box0 }
%

%

%
%
%
\unlock                         
\def\addressee#1{\line{\hskip10cm\the\date\hfill} \bigskip
   \vskip\lettertopfil
   \ialign to\hsize{\strut ##\hfil\tabskip 0pt plus \hsize \cr #1\crcr}
   \writelabel{#1}\medskip \noindent\hskip -\spaceskip \ignorespaces }
\def\signed#1{\par \penalty 9000 \bigskip \dt@pfalse
  \everycr={\noalign{\ifdt@p\vskip\signatureskip\global\dt@pfalse\fi}}
  \setbox0=\vbox{\singlespace \halign{\tabskip 0pt \strut ##\hfil\cr
   \noalign{\global\dt@ptrue}#1\crcr}}
  \line{\hskip10cm \box0\hfil} \medskip }

\def\howie{\signed{
  Sincerely Yours,\cr
  Howard E.~Haber\cr}
}



\overfullrule=0pt


%
%

\def\dslash{\not{\hbox{\kern-2pt $\partial$}}}
\def\Dslash{\not{\hbox{\kern-4pt $D$}}}
\def\Oslash{\not{\hbox{\kern-4pt $O$}}}
\def\Qslash{\not{\hbox{\kern-4pt $Q$}}}
\def\pslash{\not{\hbox{\kern-2.3pt $p$}}}
\def\kslash{\not{\hbox{\kern-2.3pt $k$}}}
\def\qslash{\not{\hbox{\kern-2.3pt $q$}}}
 \newtoks\slashfraction
 \slashfraction={.13}
 \def\slash#1{\setbox0\hbox{$ #1 $}
 \setbox0\hbox to \the\slashfraction\wd0{\hss \box0}/\box0 }


%
%

 \def\leftrightarrowfill{$\mathord-\mkern-6mu%
   \cleaders\hbox{$\mkern-2mu\mathord-\mkern-2mu$}\hfill
   \mkern-6mu\mathord\leftrightarrow$}
 \def\overlrarrow#1{\vbox{\ialign{##\crcr
       \leftrightarrowfill\crcr\noalign{\kern-1pt\nointerlineskip}
       $\hfil\displaystyle{#1}\hfil$\crcr}}}


%
%

  \def\fourptfcn#1#2#3#4#5{ \matrix{#1\cr\cr#2\cr}
      \hbox{$  \Bigg\rangle \kern -2.5pt
         {\phantom{whichcan}\over {\textstyle \downabit #5} }
           \kern-2.5pt  \Bigg\langle$} \matrix{#3\cr\cr#4\cr} }
  \def\fiveptfcn#1#2#3#4#5#6#7{ \matrix{#1\cr\cr#2\cr}
      \hbox{$  \Bigg\rangle \kern -2.5pt
        {{\pile} \lower  6pt\hbox{$\spire#5$} {\pile}\over
           \phantom{can}{\textstyle \downabit#6}\phantom{which}
             {\textstyle \downabit #7}\phantom{can}}
               \kern-2.5pt \Bigg\langle$} \matrix{#3\cr\cr#4\cr} }
  \def\spire#1{\matrix{#1\cr\Bigm|\cr}}
  \def\pile{\phantom{\matrix{1\cr2\cr3\cr}}}
  \def\downabit{\vphantom{\bigm|}}

%
%
\def\PRL#1&#2&#3&{\sl Phys.~Rev.~Lett.\ \bf #1\rm ,\ #2\ (19#3)}
\def\PRB#1&#2&#3&{\sl Phys.~Rev.\ \bf #1\rm ,\ #2\ (19#3)}
\def\PRP#1&#2&#3&{\sl Phys.~Rep.\ \bf #1\rm ,\ #2\ (19#3)}
\def\NPB#1&#2&#3&{\sl Nucl.~Phys.\ \bf #1\rm ,\ #2\ (19#3)}
\def\PL#1&#2&#3&{\sl Phys.~Lett.\ \bf #1\rm ,\ #2\ (19#3)}
\def\ZP#1&#2&#3&{\sl Z.~Phys.\ \bf #1\rm ,\ #2\ (19#3)}
\def\NIM#1&#2&#3&{\sl Nucl.~Instrum.~\& Methods\ \bf #1\rm ,\ #2\ (19#3)}
\def\RMP#1&#2&#3&{\sl Rev.~Mod.~Phys.\ \bf #1\rm ,\ #2\ (19#3)}
%
\def\NPRL#1&#2&#3&{\sl Phys.~Rev.~Lett.\ \bf #1\ \rm (19#2)\ #3}
\def\NPR#1&#2&#3&{\sl Phys.~Rev.\ \bf #1\ \rm (19#2)\ #3}
\def\NNP#1&#2&#3&{\sl Nucl.~Phys.\ \bf #1\ \rm (19#2)\ #3}
\def\NPL#1&#2&#3&{\sl Phys.~Lett.\ \bf #1\ \rm (19#2)\ #3}
\def\NZP#1&#2&#3&{\sl Z.~Phys.\ \bf #1\ \rm (19#2)\ #3}
\def\AJ#1&#2&#3&{\sl Ap.~J.\ \bf #1\ \rm (19#2)\ #3}
\def\AP#1&#2&#3&{\sl Ann.~Phys.\ \bf #1\ \rm (19#2)\ #3}
\def\NRMP#1&#2&#3&{\sl Rev.~Mod.~Phys.\ \bf #1\rm (19#2)\ #3}
%


\newbox\figbox
\newdimen\zero  \zero=0pt
\newdimen\figmove
\newdimen\figwidth
\newdimen\figheight
\newdimen\textwidth
\newtoks\figtoks
\newcount\figcounta
\newcount\figcountb
\newcount\figlines
\def\figreset{\global\figcounta=-1 \global\figcountb=-1
\global\figmove=\baselineskip
\global\figlines=1 \global\figtoks={ } }
\def\picture#1by#2:#3{\global\setbox\figbox=\vbox{\vskip #1
\hbox{\vbox{\hsize=#2 \noindent #3}}}
\global\setbox\figbox=\vbox{\kern 10pt
\hbox{\kern 10pt \box\figbox \kern 10pt }\kern 10pt}
\global\figwidth=1\wd\figbox
\global\figheight=1\ht\figbox
\global\textwidth=\hsize
\global\advance\textwidth by - \figwidth }
\def\figtoksappend{\edef\temp##1{\global\figtoks=%
{\the\figtoks ##1}}\temp}
\def\figparmsa#1{\loop \global\advance\figcounta by 1
\ifnum \figcounta < #1
\figtoksappend{ 0pt \the\hsize }
\global\advance\figlines by 1
\repeat }
\def\figparmsb#1{\loop \global\advance\figcountb by 1
\ifnum \figcountb < #1
\figtoksappend{ \the\figwidth \the\textwidth}
\global\advance\figlines by 1
\repeat }
\def\figtext#1:#2:#3{\figreset%
\figparmsa{#1}%
\figparmsb{#2}%
\multiply\figmove by #1%
\global\setbox\figbox=\vbox to 0pt{\vskip \figmove  \hbox{\box\figbox}
\vss }
\parshape=\the\figlines\the\figtoks\the\zero\the\hsize
\noindent
\rlap{\box\figbox} #3}
\def\Buildrel#1\under#2{\mathrel{\mathop{#2}\limits_{#1}}}
\def\llongrarrow{\hbox to 40pt{\rightarrowfill}}



\def\boxit#1{\vbox{\hrule\hbox{\vrule\kern3pt
\vbox{\kern3pt#1\kern3pt}\kern3pt\vrule}\hrule}}
\newdimen\str
\def\fboxit#1#2{\vbox{\hrule height #1 \hbox{\vrule width #1
\kern3pt \vbox{\kern3pt#2\kern3pt}\kern3pt \vrule width #1 }
\hrule height #1 }}
\def\fillbox#1{\hbox to #1{\vbox to #1{\vfil}\hfil}}
\def\dotbox#1{\hbox to #1{\vbox to 8pt{\vfil}\hfil $\cdots$ \hfil}}
%

%

%

%

%

%
%
%
\def\tran#1#2{\transpoint \hfuzz 5pt \gdef\label{#1}
\vbox to \the\vsize{\hsize \the\hsize #2} \par \eject }
\newdimen\baseskip
\newdimen\lskip
\lskip=\lineskip
\newdimen\transize
\newdimen\tall
\def\transpoint{\gdef\rm{\fam0\eighteenrm}
\font\twentyfourit = amti10 scaled \magstep5
\font\twentyfourrm = amr10 scaled \magstep5
\font\twentyfourbf = ambx10 scaled \magstep5
\font\twentyeightsy = amsy10 scaled \magstep5
\font\eighteenrm = amr10 scaled \magstep3
\font\eighteenb = ambx10 scaled \magstep3
\font\eighteeni = ammi10 scaled \magstep3
\font\eighteenit = amti10 scaled \magstep3
\font\eighteensl = amsl10 scaled \magstep3
\font\eighteensy = amsy10 scaled \magstep3
\font\eighteencaps = amr10 scaled \magstep3
\font\eighteenmathex = amex10 scaled \magstep3
\font\fourteenrm=amr10 scaled \magstep2
\font\fourteeni=ammi10 scaled \magstep2
\font\fourteenit = amti10 scaled \magstep2
\font\fourteensy=amsy10 scaled \magstep2
\font\fourteenmathex = amex10 scaled \magstep2
\parindent 20pt
\global\hsize = 7.0in
\global\vsize = 8.5in
\dimen\transize = \the\hsize
\dimen\tall = \the\vsize
\def\sy{\eighteensy }
\def\sl{\eighteens }
\def\bf{\eighteenb }
\def\it{\eighteenit }
\def\caps{\eighteencaps }
\textfont0=\eighteenrm \scriptfont0=\fourteenrm
\scriptscriptfont0=\twelverm
\textfont1=\eighteeni \scriptfont1=\fourteeni \scriptscriptfont1=\twelvei
\textfont2=\eighteensy \scriptfont2=\fourteensy
\scriptscriptfont2=\twelvesy
\textfont3=\eighteenmathex \scriptfont3=\eighteenmathex
\scriptscriptfont3=\eighteenmathex
\h@big=15.3\p@ \h@Big=20.7\p@ \h@bigg=26.1\p@ \h@Bigg=31.5\p@
\global\baselineskip 35pt
\global\lineskip 15pt
\global\parskip 5pt  plus 1pt minus 1pt
\global\abovedisplayskip  3pt plus 10pt minus 10pt
\global\belowdisplayskip 3pt plus 10pt minus 10pt
\def\rtitle##1{\centerline{\undertext{\twentyfourrm ##1}}}
\def\ititle##1{\centerline{\undertext{\twentyfourit ##1}}}
\def\ctitle##1{\centerline{\undertext{\caps ##1}}}
\def\vstrut{\hbox{\vrule width 0pt height 10pt depth 3pt   } \vfil}
\def\endpoint{\item{ }{\vstrut}\nextline \hangindent=0pt \hangafter=0}
\def\cline##1{\centerline{\vstrut ##1}}
\output{\shipout\vbox{\vskip 0pt
\pagecontents \vfill
\hbox to \the\hsize{\hfill{\tenbf \label} } }
\global\advance\count0 by 1 }
\rm }

%
%
\def\slacletters{\letters \let\letterhead=\SLACHEAD }
\def\ucscletters{\letters \let\letterhead=\UCSCHEAD }
\def\myucletters{\letters \let\letterhead=\MYUCHEAD }
\def\SLACHEAD{\setbox0=\vtop{\baselineskip=10pt
     \ialign{\eightrm ##\hfil\cr
        \slacbin\cr
        P.~O.~Box 4349\cr
        Stanford, CA 94309\cropen{1\jot}
        \slacphone\cr }}%
   \setbox2=\hbox{\caps Stanford Linear Accelerator Center}%
   \hrule height \z@ \kern -0.5in
   \vbox to 0pt{\vss\centerline{\seventeenrm STANFORD UNIVERSITY}}
   \vbox{} \medskip
   \line{\hbox to 0.7\hsize{\hss \lower 10pt \box2 \hfill }\hfil
         \hbox to 0.25\hsize{\box0 \hfil }}\medskip }
\def\UCSCHEAD{\setbox0=\vtop{\baselineskip=10pt
     \ialign{\eightrm ##\hfil\cr
        SCIPP\cr Natural Sciences II\cr
        University of California\cr
        Santa Cruz, CA 95064\cropen{1\jot}\cr}}
   \setbox2=\hbox{\caps Santa Cruz Institute for Particle Physics}%
   \vbox to 0pt{\vss\centerline{\seventeenrm UNIVERSITY of CALIFORNIA}}
   \vbox{} \medskip
   \line{\hbox to 0.7\hsize{\hss \lower 10pt \box2 \hfill }\hfil
         \hbox to 0.25\hsize{\box0 \hfil }}\medskip }
\def\MYUCHEAD{\rightline{\vbox{\baselineskip=10pt
         \hrule height 0pt width 0.25\hsize
         \ialign{\eightrm ##\hfil\cr
              Prof. Michael Dine\cr
              Physics Dept.\cr
              Natural Sciences II\cr
              University of California\cr
              Santa Cruz, CA  95064\cropen{1\jot}
              (408) 459--3033\cr
        Bitnet:  DINE@SLACVM\cr}}}
    \medskip }
\lock

\unlock
\def\refitem#1{\r@fitem{[#1]}}
\lock

\def\half{\ifmath{{\textstyle{1 \over 2}}}}

\def\twothirds{\ifmath{{\textstyle{2 \over 3}}}}
\def\ie{{\it i.e.}}%

\Pubnum{SCIPP 92/43}
\date{Decmeber, 1992}
\pubtype{}     
\titlepage
\vskip3cm
\title{{\bf Some Two-Loop Corrections to the Finite Temperature
Effective Potential in the Electroweak Theory}
\foot{Work supported in part by the U.S. Department of Energy.}}
\author{John E. Bagnasco and Michael Dine}
\centerline{\it Santa Cruz Institute for Particle Physics}
\centerline{\it University of California, Santa Cruz, CA 95064}
\vskip2cm
\singlespace
\vbox{
\centerline{\bf Abstract}
Perturbation theory at finite temperature suffers
from well-known infrared problems.  In the standard model,
as a result, one cannot calculate the effective potential
for arbitrarily small values of $\phi$, the Higgs
expectation value.  Because the Higgs field is now known not
to be extremely light, it is necessary to determine
whether perturbation theory is a reliable guide to properties
of the weak phase transition.  In this note, we evaluate the
most singular
contributions to the potential at two
loops as well as the leading strong interaction contributions.
Above
the critical temperature, the strong interaction corrections
are reasonably small, while the weak corrections are about 10\%,
even for rather small values of the Higgs field.
At the critical temperature, the weak corrections have a
more substantial effect, rendering the transition significantly
more first order, but not significantly changing the
upper bound on the Higgs mass required for baryogenesis.}

\vfill
\submit{Physics Letters B}
 \vfill
\endpage

\chapter{\bf Introduction}

\REF\Kirzhnitz{D.A. Kirzhnits, JETP Lett. {\bf 15} (1972) 529;
D.A. Kirzhnits  and A.D. Linde, Phys.~Lett. {\bf 42B} (1972) 471.}
\REF\shaposh{M.E. Shaposhnikov,  Phys.~Lett. {\bf 277B} (1992) 324.}
\REF\brahm{D. Brahm and S. Hsu, Caltech preprints CALT-68-1705
and CALT-68-1762 (1991).}
\REF\dhlll{M. Dine, P. Huet, R. Leigh, A. Linde and D. Linde,
Phys. Lett. {\bf B283} (1992) 319; Phys. Rev. {\bf D46} (1992) 550.}
The possibility that the observed baryon asymmetry may have been
created at the electroweak phase transition has renewed
interest in understanding the details of this transition.
The early studies of Kirzhnitz and Linde\refmark{\Kirzhnitz}
and others indicated
that in the minimal standard model, if $m_H \ll M_W$ ($m_H$ and
$M_W$ are the Higgs and $W$ mass, respectively)
this transition is
first order, becoming more weakly so as the Higgs mass increases.
This result was based on a study of the leading terms
in the finite temperature potential.  More recently, questions
were raised about this analysis; it was argued that infrared
problems might render the phase transition more\refmark{\shaposh}
or less\refmark{\brahm} weakly first order.  In particular,
it was argued that the potential, at two-loop
order, should contain linear terms
in the scalar field, $\phi$ (though of differing signs).
More careful study of the perturbation expansion, however,
shows that these linear terms cancel at this order; the phase
transition remains first order, though slightly less so\refmark{\dhlll}.

\REF\ha{G. Anderson and L. Hall, Phys.~Rev. {\bf D45} (1992) 2685.}
Still, the finite temperature theory is full of infrared problems.
As a result,  it is not possible to answer all interesting
questions in perturbation theory, even if the coupling is
weak.  The problems are easily illustrated in the minimal
standard model, with a single Higgs doublet, $\phi$.  At
high temperatures, the theory becomes essentially three-dimensional.
For massless particles, the Feynman diagrams of such
a theory exhibit power-law divergences.  In the standard model,
the gauge boson mass acts as an infrared cutoff.  This means that for
small $\phi$, the potential cannot be reliably calculated.
The results of ref.~[\dhlll] imply that perturbation theory
is good provided $M_W(\phi) \gg g^2 T$; $g$ here
denotes the gauge boson coupling.\foot{We will set the
Weinberg angle to zero in most of our discussion, since it
is not germane to the issues at hand; it is easily restored where
appropriate.}
Had the analysis of refs.~%
[\shaposh] and [\brahm] been correct, one would have had to
impose the stronger condition $M_W(\phi) \gg g T$.
We are not really interested in the behavior of the theory
for arbitrary values of the fields, however.
One question which we {\it would} like
to address in perturbation theory is the nature of the phase transition.
For this, one needs to study the behavior of the potential
near the minimum.
As we will review below, for weak coupling in this theory
the minimum of the potential occurs for $g  \phi \propto g^4T/\lambda$,
where $\lambda$ is the quartic Higgs coupling.  Thus for sufficiently
small $\lambda$, the perturbation expansion should be reliable.
Indeed, the expansion becomes an expansion in powers of
$g^2$, $\lambda$, and $\lambda/g^2$.
If perturbation theory is to be a useful guide to the behavior of this
system, {\it all} of these quantities must be small.

\REF\higgslimit{M.E. Shaposhnikov, JETP Lett. {\bf 44} (1986) 465;
Nucl.~Phys. {\bf 287} (1987) 757; Nucl.~Phys. {\bf 299} (1988) 797;
A.I. Bochkarev, S. Yu. Khlebnikov and M.E. Shaposhnikov,
Nucl.~Phys. {\bf B329} (1990) 490.}
On the other
hand, we  already know from LEP that the Higgs particle cannot be
extremely
light, and so the quartic coupling cannot be arbitrarily small.
Even
in extensions of the standard model, designed to produce
a more strongly first order transition\refmark{\ha}
there are likely to be important constraints on the smallness of
various couplings.
Moreover, at finite temperature, one does not necessarily obtain
the factors of $4 \pi$ familiar
in zero temperature calculations, and corrections are potentially
large.  It is thus natural to ask whether perturbation theory
is a good guide in the experimentally allowed regions of parameter
space.

In this note, we investigate some of these questions at two-loop
order in the minimal standard model.
It is known that in this
model one cannot produce an asymmetry,
given the present bound on the Higgs mass\refmark{\dhlll,\higgslimit}.
We choose the minimal model
to illustrate our points because of its simplicity; our
analysis is readily extended (often trivially extended)
to other theories in which the phase transition is likely
to be more strongly first order.  A complete evaluation of all
two-loop diagrams is
quite involved.  However, it turns out that there are certain
leading contributions which can be easily computed.
The simplest to describe are the QCD corrections.  These are potentially
important for top quark loops, due to the large value of the top quark
mass and of the strong gauge coupling.
They can be obtained
by suitably adapting existing QED computations.
In addition,
there are certain purely electroweak effects which are enhanced
by powers of logarithms of the scalar field.  We have already remarked
that in ref.~[\dhlll] it was shown that certain leading
infrared effects cancel.  It turns out
that the subleading terms come with
logarithms of the gauge boson mass over the temperature.
We will compute a set of terms in the potential
proportional to $g^4\phi^2 \log({g\phi/2T})$.
We will see that, away from the critical temperature,
$T_c$, for the phase transition,
both the QCD and pure electroweak effects
give corrections of order a few per cent to the potential.
Thus perturbation theory appears to be in quite good shape.
Near $T_c$, however, the situation is more delicate;
the weak interaction corrections are in fact quite large,
here, and seem to make the transition more first order.

\chapter{Review of Earlier Results}

\REF\lindebook{A..D. Linde, {\it Particle Physics and Inflationary
Cosmology} (Harwood, New York, 1990).}
\REF\sher{M. Sher, Phys.~Rep. {\bf 179} (1989) 273.}
The zero temperature potential, taking into account
one-loop corrections, is given by\refmark{\lindebook,\sher}
$$
V_0 = - {\mu^2\over 2}\phi^2 + {\lambda\over 4} \phi^4 +
2Bv_o^2\phi^2 - {3\over 2} B\phi^4 + B \phi^4 \ln\left({\phi^2\over
v_o^2}\right) \ .
\eqn\zerotemp$$
Here
$$
B = {3\over 64 \pi^2 v_o^4} (2 m_W^4 + m_Z^4 - 4 m_t^4) \ ,
\eqn\bdefined$$
$v_o = 246$ GeV is the value of the scalar field at the minimum
of $V_0$, $\lambda = \mu^2/v_o^2$, $m^2_H = 2\mu^2$.
At a finite temperature, one should add to this expression the term
$$
V_T = {T^4\over 2 \pi^2} \left[ 6I_{-}(y_W) + 3I_{-}(y_Z) -
12I_{+}(y_t)\right] \ ,
\eqn\tempcorrection$$
where $y_i = M_i\phi/v_o T$, and
$$
I_{\mp}(y) = \pm \int_{0}^{\infty} dx
\ x^2 \ln \left(1  \mp e^{- \sqrt{x^2+y^2}}\right) \ .
\eqn\iminus$$
In the high temperature limit it is sufficient to use an
approximate expression for $V(\phi,T)$\refmark{\Kirzhnitz,\ha}
$$
V(\phi,T) = D (T^2 - T_o^2) \phi^2 - E T \phi^3 +
{\lambda_T\over 4} \phi^4 \ .
\eqn\hightemp$$
Here
$$
D = {1\over 8v_o^2} ( 2 m_W^2 + m_Z^2 + 2 m_t^2) \ ,
\eqn\dequals$$
$$
E =  {1\over 4\pi v_o^3} ( 2 m_W^3 + m_Z^3) \sim 10^{-2} \ ,
\eqn\eequals$$
$$
T^2_o = {1\over 2D}(\mu^2 - 4Bv_o^2) =
{1\over 4D}(m_H^2 - 8Bv_o^2) \ ,
\eqn\tzero$$
$$
\lambda_T = \lambda - {3\over 16 \pi^2 v_o^4}
\left( 2 m_W^4 \ln{m^2_W\over a_B T^2} +
m_Z^4 \ln{m^2_Z\over a_B T^2} -
4 m_t^4 \ln{m^2_t\over a_F T^2}\right) \ ,
\eqn\lambdat$$
where $\ln a_B = 2 \ln 4\pi - 2\gamma \simeq 3.91$,
$\ln  a_F = 2 \ln \pi - 2\gamma \simeq 1.14$.
\REF\carring{M.E. Carrington, Phys. Rev. {\bf D45} (1992) 2933.}
As explained in refs.%
[\dhlll] and [\carring],
the constant $E$ is
reduced by higher order corrections by a factor of $2/3$.

This potential leads to a  phase transition which is at least weakly
first order, basically as a consequence of the term cubic in $\phi$.
At very high temperatures, $V$ has a unique minimum at $\phi=0$.
As the temperature is decreased, a second minimum appears
at a temperature
$$T_1^2 = {T_o^2 \over 1 - 9 E^2/8 \lambda_{T_{1}} D}.\eqn\tone$$
At a temperature $T_c$, this new minimum becomes degenerate
with the minimum at the origin; $T_c$ and the corresponding minimum
of the potential $\phi_c$ are given by
$$T_c^2={T_o^2 \over 1 - E^2/\lambda_{T_{c}} D}~~~~~~~~~~{\phi_c ={2E T_c
\over \lambda_{T_c}}}. \eqn\tc$$ Finally,
at $T_o$, the minimum at the origin disappears, and the potential has
a unique minimum.

\REF\dhss{M. Dine, P. Huet, R. Singleton and L. Susskind, Phys.~Lett.
{\bf 257B} (1991) 351.}
\REF\mvst{L. McLerran, M. Shaposhnikov N. Turok and M. Voloshin,
Phys.~Lett. {\bf 256B} (1991) 451.}
\REF\bubble{A.D. Linde, Phys.~Lett. {\bf 70B} (1977) 306;
{\bf 100B} (1981) 37; Nucl.~Phys. {\bf B216} (1983) 421.}

\FIG\topquark{Two-loop contribution to the potential involving
a gluon correction to a top quark loop.}

\chapter{QCD Corrections}
\REF\kapusta{J.I. Kapusta, Nucl. Phys. {\bf B148} (1979) 461.}
\REF\kbook{J.I. Kapusta, {\it Finite Temperature Field Theory},
(Cambridge University Press, Cambridge, 1989).}

Because they are conceptually simplest, we consider, first, the strong
corrections to the potential.  The leading terms arise from a gluonic
correction to a top quark loop (fig.~\topquark.)  Apart from a trivial
group theory factor, this calculation is identical to the calculation
of the two-loop free energy in QED\refmark{\kapusta,\kbook}.

\topinsert
   \tenpoint \baselineskip=12pt   \narrower
\vskip12pt\noindent
{\bf Fig.~\topquark.}\enskip
Two-loop contribution to the potential involving
a gluon correction to a top quark loop.
\vskip12pt
\endinsert

If one examines the Feynman diagram containing the top quark (electron
in QED) they are completely free of infrared problems even as the
quark (electron) mass tends to zero.  This is because the discrete
Matsubara frequencies for fermions are all non-zero.  However, the
results of the QED calculation\refmark{\kapusta,%
\kbook} diverge logarithmically with the mass [as
$m_e^2 \ln(m_e)$].  The reason is not difficult to understand.  The QED
calculation is performed by subtracting the electron mass on shell.
The required counterterm itself diverges logarithmically.  In QCD, we
want to subtract the mass at some off-shell point; in the present
case, we are interested in the mass (or, more precisely, $g_t$), at a
scale of order $T$, the temperature.   Renormalizing
off mass shell in this way, there are no logarithms.

To evaluate the top quark contribution, we begin with the QED
expression (for zero chemical potential)\refmark{\kapusta,\kbook} $$V
=  {1 \over 3} e^2 T^2 \int{d^3 p \over (2 \pi)^3}{n_F(p) \over E_p}
+e^2 \int{d^3 p \over 2 \pi^3} {d^3 q \over 2 \pi^3}{1\over E_p E_q}$$
$$
\times~ 2 \left[1~+~ { m^2 \over (E_p -E_q)^2 - (\vec p -\vec q)^2}~+~
 { m^2 \over (E_p +E_q)^2 - (\vec p -\vec q)^2}\right]n_F(p)n_F(q)
\eqn\qed$$
where $$E^2 = \vec p^{~2} + m^2 ~~~~~~~~~~~n_F(p) = {1 \over e^{\beta
E_p} + 1}\ .$$

\REF\dolan{L. Dolan and R. Jackiw, Phys. Rev. {\bf D9} (1974) 3320.}
We will content ourselves with extracting the term of order $m^2$.
The terms involving involving $m^2$ explicitly
in eq.~\qed\ must be handled with some care (we thank Peter Arnold for
pointing this out to us).  It is necessary to perform the angular
integrals first
before taking the $m\rightarrow 0$ limit.  The integration
is then straightforward, apart from the integral
$$\int dxdy\ln{(x-y)^2 \over (x+y)^2 }{1 \over (1+e^x)(1+e^y)}
={\pi^2 \over 3} (\ln(2)-1)\eqn\howie$$
(We thank Howard Haber for evaluation of this integral.)
The remaining
integrals factorize, and are readily evaluated using
standard tricks (\eg, see appendix A of ref.~[\kbook]).  In particular,
the integral
$$f_3(y) = {1 \over 2} \int^{\infty}_0 {x^2 dx \over (x^2
+ y^2)^{1/2}}~ {1 \over e^{\sqrt{x^2 + y^2}}+1} \eqn\standardintegral$$
is given by
$$f_3 = {\pi^2 \over 24} + {y^2 \over 8} \left( \ln y +
\gamma - \ln\pi -{\half} \right)\ . \eqn\fthree$$

As we have noted above, this does not yet give us the result we want;
we need to add a finite counterterm to change the result from an
on-mass-shell subtraction to an off-shell one.  We choose to use the
$\overline{MS}$ scheme.  A straightforward calculation gives that in
the {\it one-loop} result, one must replace $m$ by
$$m \rightarrow m\left\{1- {3e^2
\over 16 \pi^2} \left[ \ln\left({m^2 \over \mu^2}\right)
- {4 \over 3}\right]\right\}\ .
\eqn\massrenorm$$
To order $m^2$, the lowest order result is $m^2/ T^2$.
Putting everything together
then gives for the fermionic contribution
$$V_f= {-7 \pi^2 T^4\over 60 } + {m^2 T^2\over 4} +
{5g^2 T^4\over 72} + {g^2 m^2 T^2\over 4 \pi^2}
\left[-\half - \ln(\pi) +\gamma+\twothirds \ln(2)\right]\ .
\eqn\topresult$$
The corrections both to the leading $T^4$ term and to the
quadratic term are about $8\%$.

\chapter{Pure Electroweak Corrections}

We now turn to the purely electroweak corrections.
It is important, first, to understand the nature of the perturbation
expansion at high temperatures.  In general, if
one examines the Feynman diagrams for the potential, one sees
that they diverge in the infrared as the gauge boson masses
($\sim {g \phi/2 }$) tend to zero.  The divergences are cut off by
the gauge boson masses themselves.  For example, at one-loop,
the coefficient of the quartic term in the potential is linearly
divergent as the mass tends to zero; this is the origin of the cubic
term in the potential.
At higher loops one encounters more and more severe divergences.
{}From the start, then, it is clear that
perturbation theory cannot be good for arbitrarily small values of
$\phi$.  One can hope, however, that if $\lambda$ is small enough, the
expectation value of the scalar field will be large enough that
perturbation theory will be valid near the minimum of the potential.
Note that the one-loop potential gives for the location of the minimum
at $T_c$ a result which goes as $g^3/\lambda$.  The potential
at the minimum is of order $g^{12} /\lambda^3$.

The analysis of refs.~[\shaposh] and [\brahm]
suggested that at two-loop order the
coefficient of the quadratic term in the potential is linearly
divergent.  Extended to higher orders, this analysis would
suggest that the loop expansion parameter is $g^2 /M_W^2$;
for $\phi \propto  g^3/\lambda$, this gives $\lambda^2/g^6$
at the minimum.  In ref.~[\dhlll], it was shown that there are
no such linear terms; a simple extension of the arguments
given there shows that the expansion parameter is $g^2 /M_W$,
or $\lambda/g^2$ at the minimum.

\FIG\dangerous{Two-loop diagrams which potentially give
linear terms in the potential.}
\FIG\reorganized{The two diagrams of fig.~\dangerous\ can be
viewed as a correction to the one-loop gauge boson propagator.}
\REF\iz{C. Itzykson and J.C. Zuber,  {\it Quantum Field Theory,}
(McGraw Hill, New York, 1980).}
\REF\muta{T. Muta, {\it Foundations of Quantum Chromodynamics},
(World Scientific, Singapore, 1987).}

\topinsert
   \tenpoint \baselineskip=12pt   \narrower
\vskip12pt\centerline{%
{\bf Fig.~\dangerous.}\enskip
Two-loop diagrams which potentially give
linear terms in the potential.}
\vskip24pt
\vskip12pt\noindent
{\bf Fig.~\reorganized.}\enskip
The two diagrams of fig.~\dangerous\ can be
viewed as a correction to the one-loop gauge boson propagator.
\endinsert

Actually, while it was shown in ref.~[\dhlll] that the leading
divergences cancel, the subleading terms come with logarithms
of the gauge boson mass, \ie, with
$\ln(\phi^2)$.  To understand
how this comes about, it is helpful to describe the results
of ref.~[\dhlll] in a slightly different fashion than
explained in that paper.  We assume that the Higgs mass is small
and neglect Higgs self-interactions.
We work, as there, in Coulomb gauge;
the theory in the infrared is then like a three-dimensional
theory in Landau gauge.  As explained there, for small $\phi$
one can ignore the contributions of the longitudinal modes.
The diagrams involving transverse gauge boson loops
of fig.~\dangerous\ are each
potentially quite singular in the infrared.  Indeed,
while each diagram is formally of order $\phi^2$, power-counting
suggests that the integrals separately behave as $1 /\phi$,
and that these diagrams should thus give linear terms in $\phi$
(this is the source of the linear terms in
refs.~[\shaposh] and [\brahm]).
However, the leading divergences cancel between the two diagrams.
This is easily understood; the diagrams can be organized as in
fig.~\reorganized.
In other words, one can view the combination of diagrams as a
propagator correction to the one-loop gauge boson contribution
to the free energy.  To verify this requires care with the combinatorics;
as explained in ref.~[\dhlll], it is simplest to verify statements of
this kind for tadpoles, rather than directly for the free energy.
In any case, it is well-known that the three-dimensional gauge
boson propagator vanishes at zero momentum as a consequence of
gauge invariance.  A straightforward calculation
\foot{This calculation can be found in many textbooks;
see, \eg, refs.~[\iz-\muta].  However,
note that the formula the first reference contains a typographical
error; we have verified that the formula in the second is correct.}
gives
$$\Pi_{ij}={11g^2T \over 32 \sqrt{q^2}}(q^2\delta_{ij} - q_{i}
q_{j})\ . \eqn\propagator$$
Because the polarization tensor vanishes linearly with the
momentum, the integral behaves for small $\phi$ as
$\ln(\phi)$.  A simple calculation gives
$$ \delta V_{1} =-{33g^4 T^2 \phi^2\over 256 \pi^2}\ln(g\phi/2T)\ ,
\eqn\leadinglog$$
(note that, as is conventional in the finite temperature
theory, we are again normalizing $\phi$ as a real field here).

\FIG\corrections{}
There are other corrections at two loop order which depend on
$\log(\phi)$.  These come from the diagrams in fig.~\corrections,
which contain scalar fields in intermediate states.  These are
somewhat more subtle to treat, since one must perform a resummation
of some type in order to determine the effective scalar propagator.
However, without probing this issue too deeply, it seems likely
that any resummed scalar mass will be smaller than the gauge boson mass.
For example, even well above the phase transition, the effective
scalar mass in the one loop potential is somewhat smaller than
the gauge boson mass; near the transition, it is much smaller.
So to get an estimate of these effects, we will simply neglect it.
In this case, the diagrams are straightforward to evaluate.\foot{In
evaluating the first diagram, one approach is to route the loop momenta,
$\vec l$ and $\vec k$ so that the scalar propagator depends on $\vec k-
\vec l$.  This propagator can then be expanded in a (double) series
of Legendre polynomials.  The momentum integrals become trivial,
and the sums give $\zeta$-functions.}
{}From the first, we obtain
$$ \delta V_{2} = {9g^4 T^2 \phi^2\over 512 \pi^2}\ln(g\phi/2T)\ ,
\eqn\leadinglogb$$
while from the second  we have
$$ \delta V_{3} ={3g^4 T^2 \phi^2\over 256 \pi^2}\ln(g\phi/2T)\ .
\eqn\leadinglogc$$
So overall, the shift is
$$ \delta V_{1} =-{51 g^4 T^2 \phi^2\over 512 \pi^2}\ln(g\phi/2T)\ .
\eqn\leadinglogtot$$

\topinsert
   \tenpoint \baselineskip=12pt   \narrower
\vskip12pt\centerline{%
{\bf Fig.~\corrections.}\enskip
Diagrams with scalers in intermediate states.}
\endinsert

\FIG\potentials{One-loop potential at $T_c$ and two-loop potential
at $\widetilde T_c$ for a 35~GeV Higgs.}
Away from the critical temperature, this correction is not too
large.  If the logarithm is of order $2$ (corresponding to
typical $\phi$'s at the transition), we only obtain a
$10\%$ correction.  However, near the phase transition, the
effect of this term is enhanced, because the quadratic term
is so small.  For example, keeping only the one-loop correction,
the coefficient of the quadratic term at $T_c$ is
of order $10^{-2}T_c^2$.  This is smaller than the size of the
logarithmic term.
So the effect of the logarithmic
correction is potentially dramatic.  In order to evaluate its
significance, note first that the various two-loop corrections
we have considered alter the value of the critical temperature.
Taking $m_H=35$~GeV, we have first found this new temperature,
$\widetilde T_c$
In fig.~\potentials,
we have presented two curves:  the one-loop result (with the
correction of ref.~[\dhlll])
at $T_c$, and the two-loop
result, at $\widetilde T_c$.  The two-loop correction is seen to
render the transition more first order; at the peak,
the correction is almost a factor of $3$.  One can
see in fig.~\potentials\ that the location of the
minimum changes by about $20\%$.  This weakens somewhat
(by about 4 or 5 GeV) the
bounds mentioned earlier on the Higgs mass.

It is reasonable to ask about the gauge-dependence of this result.
This is a subtle question which we will not explore here.  We expect,
however, that for physically meaningful questions (such as the
sphaleron rate at the minimum) these results should be gauge-independent.

\topinsert
   \tenpoint \baselineskip=12pt   \narrower
\vskip10.5cm\centerline{
{\bf Fig.~\potentials.}\enskip
One-loop potential at $T_c$ and two-loop potential
at $\widetilde T_c$ for a 35~GeV Higgs.}
\vskip12pt
\endinsert

\chapter{Conclusions}

{}From this work, it appears that perturbation theory at finite temperature
in the standard model is not in terribly bad shape, even for Higgs
fields which are not extremely light.  Away from the critical temperature,
perturbative corrections are quite small.  Near $T_c$, however,
the situation is somewhat more subtle.  Because the  leading
quadratic terms in the potential vanish at this point, subleading
terms which are not precisely quadratic (\ie, have logarithmic
modifications) are of great importance.  At two-loop order,
in the minimal standard model,
they render the transition significantly more first order,
but only slightly modify the limits on the Higgs mass.

It is natural to ask about the effects of still higher order
corrections.  Examination of some particular graphs leads us to
suspect that the picture presented
here will not be drastically altered.  Still, given the
size of the corrections which we have found, non-perturbative
calculations would be of great interest.  In particular, we
have not carefully investigated the problem of corrections to the
potential loops with Higgs fields only, which might be quite important at the
transition.

\bigskip\noindent
{\bf Acknowledgements:}

We thank P. Huet, R. Leigh, and A. Linde for discussions and suggestions.
As we were completing this work, we received a paper by Peter
Arnold and Olivier Espinosa on the two loop corrections at finite
temperature.  These authors have thoroughly analyzed the
perturbative corrections to the potential, including carefully
the non-logarithmic corrections, as well as corrections involving
longitudinal gauge bosons.  They indeed find that the largest
corrections are the logarithmic terms found here.  We are grateful
to them for pointing out an error in our earlier computation of the
QCD corrections, and to Howard Haber for valiantly performing a
challenging integral.

\bigskip
\refout
\end